\documentclass[letter]{aa}    

\usepackage{graphicx}
\usepackage{txfonts}
\begin{document} 

\title{The neutral gas phase nearest to supermassive black holes}
\subtitle{A massive neutral-atom- and molecule-rich  broad line region in active galactic nuclei}
   \author{Thi W.-F.  \inst{1}
          \and
          Papadopoulos, P. P. \inst{2,3,4}
          }
        \institute{ 
        Max Planck Institute for Extraterrestrial Physics, Giessenbachstrasse, 85741 Garching, Germany \email{wingfai.thi@googlemail.com}
        \and
         Department of Physics, Section of Astrophysics, Astronomy and Mechanics, Aristotle University of Thessaloniki, GR-54124 Thessaloniki, Greece 
              \email{padelis@auth.gr}
         \and
         Research Center for Astronomy, Academy of Athens, Soranou Efesiou 4, GR-11527 Athens, Greece
         \and
          School of Physics and Astronomy, Cardiff University, Queens Buildings, The Parade, Cardiff CF24 3AA, UK
              }

   \date{... ; ....}
 
  \abstract
      {Broad line regions (BLRs)  are known to contain gravitationally bound gas   within a
        r$\sim $(few) $\times$ ($10^2$-$10^3$) Schwarszchild
radii ($\rm R_{S}$) near supermassive black holes
        (SMBHs) in active galactic nuclei (AGNs). Photo-ionized by a strong non-stellar AGN continuum, this gas emits
        luminous ultraviolet/optical/near-infrared lines from ionized hydrogen (and other multi-ionized atoms) that have the widest velocity profiles observed in galaxies, uniquely indicating the deep gravitational wells of SMBHs.}
      {Nearly all BLR studies focus on its ionized gas phase  (hereafter BLR$^{+}$), with typical masses of
        only $\sim $(few) $\times$ (10-100)\, $\rm M_{\odot}$, despite strong indications
        of  neutral BLR gas reservoirs (hereafter, BLR$^{0}$) with $\rm M_{BLR^{0}}$$\sim $$10^{5-6}$\,$\rm M_{\odot}$.}
      { We used the photoionization code {\sc CLOUDY}, with its chemistry augmented using three-body reactions, to explore 1D  models of  dustless BLRs,
        focusing on  the BLR$^{0}$ conditions and the abundances of its  most prevalent neutral atoms and molecules.}
      { A (neutral-atom-) and molecule-rich BLR$^{0}$ gas phase is found to be underlying the BLR$^{+}$. The latter 
        occupies only a thin outer layer of AGN-irradiated gas column densities, while the former  contains the bulk of the  BLR  gas mass.  Atomic carbon and oxygen as well as the CO molecule  can  reach substantial abundances in
        the BLR$^{0}$, while their lines at infrared (IR) and submillimeter (submm) wavelengths  can yield  new probes of the
        BLR physical conditions and dynamics, unhindered by the dust absorption from outer AGN tori that
        readily absorb the BLR$^{+}$ optical and far-ultraviolet (FUV) lines.}
      {We find that neutral-atom-rich and even molecule-rich gas  can exist in the BLR$^{0}$. The corresponding spectral lines from
        neutral atoms and molecules promise a new spectral window of gas dynamics  in the vicinity of~SMBHs unhindered
        by dust absorption. This may even offer the prospect of conducting novel tests of general relativity in strongly curved spacetime.}

 \maketitle
\titlerunning{Neutral-atomic/molecular-rich gas near SMBHs}
\authorrunning{Thi \& Papadopoulos}
   
%

\section{Introduction}\label{introduction}

The probability of  a neutral-atomic or even  molecule-rich gas phase
having any significant  mass lying in the  vicinity ($\sim $$\rm{(few} \times
$($10^2$--$10^3$)$R_{\rm S}$), $R_{\rm S} = 2GM_{\mathrm{SMBH}}/c^2$)  of
a supermassive black  holes (SMBH)  in active  galactic nuclei  (AGNs)
could be naively be considered as being  virtually zero, given the extreme ultraviolet (UV)
and hard X-ray non-stellar continua  emanating from the SMBH accretion
disks and  their  hot  coronae,  (e.g., \citealt{Rok92,  Thom16,
  Dew21}).   The  strong  UV  AGN  continuum  photoionizes  such  gas,
sublimating its concomitant dust once  its temperature rises above the
sublimation  limit of $T_{\rm dust}$$\sim  $(1500-2000)\,K; this limit can be reached  at
distances   of   $r \leq R_{\rm sub} \sim 0.2[L_{\rm bol}/(10^{46}\rm erg\,s^{-1})]^{1/2}$  
 pc from   the    AGN   (e.g.,
\citealt{Bas18}). Photoionized gas seems an  unlikely phase for the onset of
a  (neutral-atom and molecule)-rich  chemistry; even  a  neutral  but
dustless gas phase would also seem to be as well, given the role of dust in catalysing
$\rm H_2$ formation (\citealt{Gou63, Ca02, Thi2020}, and references therein).\\
Three important  elements are  missing from this simple picture, namely: 
a) a molecule-rich  chemistry is possible even in
hot    dustless   gas    provided    densities    are   high    enough
($n_{\rm H}\geq 10^{10}$\,cm$^{-3}$); b)  molecules such as $\rm H_2$  or CO can strongly
self-shield once formed even in  initially small quantities; and c) the
thermal  (and thus  chemical) gas  properties are  drastically altered
once neutral atomic  and/or molecular species form,  even in initially
low abundances because they are efficient gas coolants via their lines.
Such a molecule-rich  chemistry in  hot, strongly UV-irradiated  dense gas,
whose concomitant dust reservoir is fully sublimated, has already been
inferred in the disk around  the B star \object{51 Oph} \citep{Thi05, Thi13}. This result has subsequently been 
confirmed    by   spatially  resolved   near-infrared (NIR)   interferometric
observations of  CO overtone lines \citep{Tat08, grav5}. In addition, CO
has  even  been  detected  in the  X-ray-irradiated  gas  ejecta  of
SN\,1987A   \citet{spy88},  its   formation   mediated  by   He$^{+}$
charge-transfer reactions  \citep{Lep90} in  an environment  that
certainly  ranks among the  most hostile  ones in  the Universe  with respect to a
molecule-rich~chemistry.\\

The possibility of a neutral-atom- and molecule-rich gas phase in the vicinity
of SMBHs,  despite photoionization  by a
strong  UV/X-ray AGN continuum,  has been  investigated  in the past
(\citet{Lep85}); however, these studies have not   included   the  very
important process  of dust  sublimation by  the AGN  radiation fields.
This work  was revisited by \citet{Kal87}, which did not include H$^{-}$ electron photo-detachment by IR radiation,
another process that is fundamentally important for the chemistry in the strongly AGN-irradiated BLR gas.
\citet{fer89}, modeled the \ion{Ca}{II} line triplet AGN observations, 
 thereby correcting the two aforementioned omissions and arguing  for the presence of a massive 
neutral BLR  gas  phase underlying the photo-ionized one. However these authors 
did not consider any neutral-atomic or molecular abundances past the 
BLR$^{+}$/BLR$^{0}$ boundary\footnote{They  made  a comment only about
 H$_2$ (in their Appendix), which unlike \citet{Kal87}, found it to be
 only a trace species, but otherwise giving no abundance estimates
 or profiles.},   reporting $N(\rm{H})_{\rm{min},0}$ $\sim$ $10^{24.5}$\,cm$^{-2}$ 
 as the minimum column density necessary  for explaining the \ion{Ca}{II} triplet IR-line strengths 
 relative to its UV lines and other lines emanating from the BLR$^{+}$.
 Nevertheless, the very same arguments presented by \citet{fer89}, along with the obvious
 expectation that neutral-atoms and molecules will form at the deeper end of any
 1-side irradiated BLR gas column densities,  (e.g., \citealt{Gas09}), as well as general arguments 
 made throughout the AGN literature about an underlying BLR$^{0}$ being much more massive
 than the BLR$^{+}$, (e.g., \citealt{Bal03, fer04}), all require any new state-of-the-art
 BLR models to extend well past the aforementioned minimum gas column density.

In this work, we examine the thermal state and the abundances of a few key
 atomic and molecular species in a  dustless BLR$^{0}$, for gas columns 
  up to $\sim $(few)$\times $$10^{26}$\,cm$^{-2}$, and
 a typical AGN SED and radiation  field strength. We then briefly
 comment on the possible use of neutral-atom and/or molecular  line emission at
 IR and submillimeter (submm) wavelengths as extinction-free probes of the bulk of the BLR
  mass and the gas velocity fields  near  SMBHs. It may even be useful for carrying out novel tests
 of general relativity (GR) near SMBHs,  as recently proposed by \citet{kos24}.

\section{The {\sc CLOUDY} code}\label{cloudy}

We utilized the {\sc CLOUDY 23.01} \citep{Cloudy2023} photoionization and photodissociation code with the standard AGN-input spectral energy distribution (SED). {\sc CLOUDY} includes photodissociation region (PDR) physics and has been benchmarked against other (PDR) codes  \citep{Rollig2007}. We adopted a one-sided illuminated plan-parallel geometry. The chemical network is based on the UMIST 2012 Database for Astrochemistry, with enhanced metal abundance ($Z/Z_\odot=5$) and specific extensions of the chemical rates \citep{Shaw2022}; these are especially relevant for  chemical reaction rates at $T_{\mathrm {gas}} \geq $100\,K \citep{Shaw2023RNAAS}.
Reactions relevant to PDR modeling are present in the {\sc CLOUDY} network, including reactions with excited H$_2$. The default network has been complemented by relevant three-body and calcium reactions (see Appendices \ref{Appendix_chem} and \ref{calcium_network}).
The relative abundances follow the default ISM values. No grains nor PAHs are present because their temperature would be higher than their sublimation temperature in most part of the BLR, although the gas and dust temperature can be below the dust sublimation limit at very high column densities. The potential effects of any remaining dust grains in the BLR will be further explored in a forthcoming paper. In the absence of dust grains, H$_2$, C,  and CO can only self-shield and self each other against photodissociation \citep{Hollenbach1999}. The cosmic-ray flux density was set to the default value. 

\section{The BLR$^{0}$ vs the BLR$^{+}$: Model parameters}\label{parameters}

Ever since their discovery \citep{seyf}, the luminous and very broad
($FWZI$$\sim$($10^3 - 10^4$)\,km\,s$^{-1}$) lines  of the  BLR$^{+}$ at
UV/optical/NIR wavelengths have been AGN signposts, revealing the
deep gravitational potential of SMBHs within $\sim $(500-$10^4$)\,$
R_S$.   Powered  by photoionization  from  a  strong non-stellar  AGN
continuum  that yields  ionized hydrogen  (and other  multiply ionized
atoms),  these  lines are  used  to  study  the BLR  conditions  (e.g.,
density,      ionization,       temperature,      and      metallicity
\citealt{Netz90,Kor04,Pet06,Gas09}).   However  the   sharpness  of  the
H$^{+}$/H transition zone with

\begin{equation}
\frac{\Delta R}{R_{+}}\sim 3.5\times 10^{-4}\left[\left(\frac{S_\mathrm{ion}}{10^{49}\,\mathrm{s}^{-1}}\right)\left(\frac{n}{100\,\mathrm{cm}^{-3}}\right)\right]^{-1/3}
\end{equation}

\noindent
giving the  width $\Delta R$  of this zone  (assuming a
Str\"{o}mgren  sphere  of  photoionized gas  of radius,  
$R_{+}$,  with  $S_{\mathrm{ion}}$ as the  ionizing  source's  photoionization
rate). This renders  the BLR$^{+}$ main lines (i.e., the H recombination
lines) incapable  of probing  the BLR$^{0}$ and, thus, of the total
  (BLR$^{0}$+BLR$^{+}$)    gas      mass as well. For instance, this gives
$n_{\rm H}=10^{11}$\,cm$^{-3}$:   $\Delta   R/R_{+}$$\sim   $$7.5\times
10^{-7}$$[S_{\rm ion}/(10^{49}\,\mathrm{s}^{-1})]^{-1/3}$$\ll  1$ for  typical
$S_{\mathrm{ion}}$  rates  for  AGNs.    Lines  from  ionized  and neutral atoms 
with ionization  potentials lower  than that  of H  (e.g., Fe$^{+}$,
Ca$^{+}$, C, and O) is able  probe the neutral gas past the H$^{+}$/H  transition zone,
yet  deeper within, some can be locked into molecules, even in the absence of dust.

This leaves two key aspects open for inquiry: a) the actual value BLR$^{0}$ gas mass and
b)  what lines are able to  probe it  in its entirety. We must first adopt plausible masses
for the BLR$^{0}$ in order to set an $N(\rm{H})_{max}$ value in our 1-D {\sc CLOUDY} models.
This is a critical parameter since, for  any given  AGN  radiation  field and  SED  incident  on a 
one-sided irradiated BLR  gas layer, arbitrarily  large BLR$^{0}$ columns  can be
obtained  beyond  the H$^{+}$/H  transition  zone,  for a high  enough
$N(\rm{H})_{\rm max}$ result. Direct BLR$^{+}$ gas mass estimates via hydrogen  recombination lines
give $M_{\mathrm{BLR}^{+}}$$\sim $(10--100)\,M$_{\odot}$, while a set
of necessarily more  indirect arguments set the  putative BLR$^{0}$ to
be at   least   $M_{\mathrm{BLR}^{0}}$$\sim   $($10^4$--$10^5$)\,
M$_{\odot}$ \citep{Bal03,  Gas09, fer04}.  On the other hand an $N(\mathrm{H)_{min}}$
value for BLR  gas structures can be  obtained by simply demanding  the SMBH
gravity dominates over  the radiative pressure from  the AGN radiation
field (i.e., the BLR is gravitationally bound to the SMBH). This yields:
\begin{equation}
\langle N(\mathrm{H}) \rangle >  \frac{L_{\mathrm{AGN}}}{4\mathrm{\pi} G m_{\mathrm{p}} c M_{\mathrm{SMBH}}},
\end{equation}
\begin{figure*}[!h]
\centering
\includegraphics[angle=0,width=9cm,height=6cm,clip]{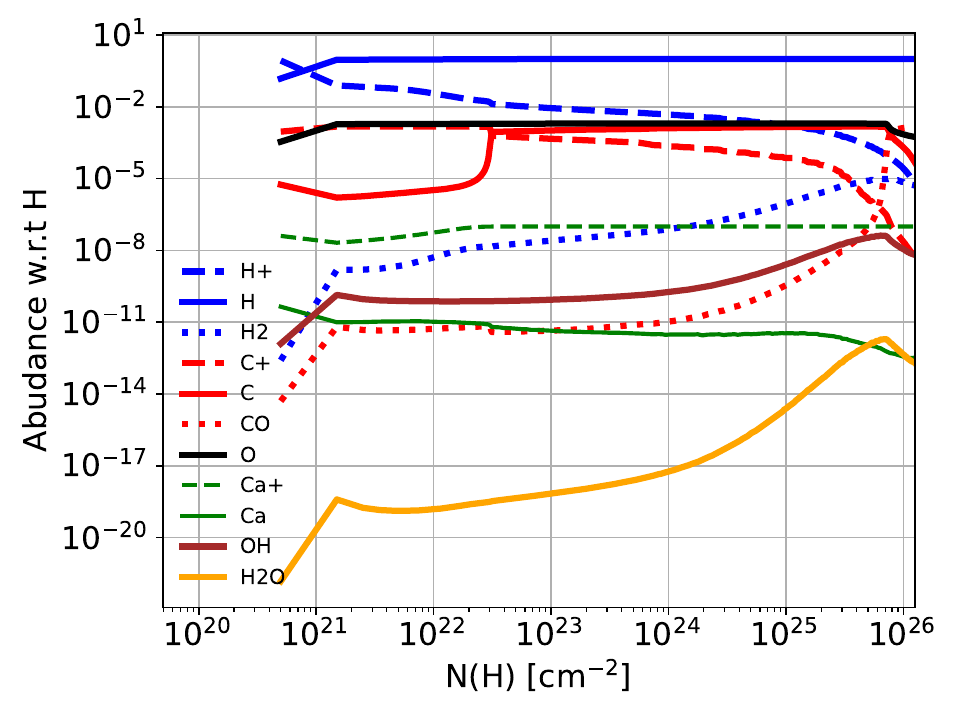}
\includegraphics[angle=0,width=9cm,height=6cm,clip]{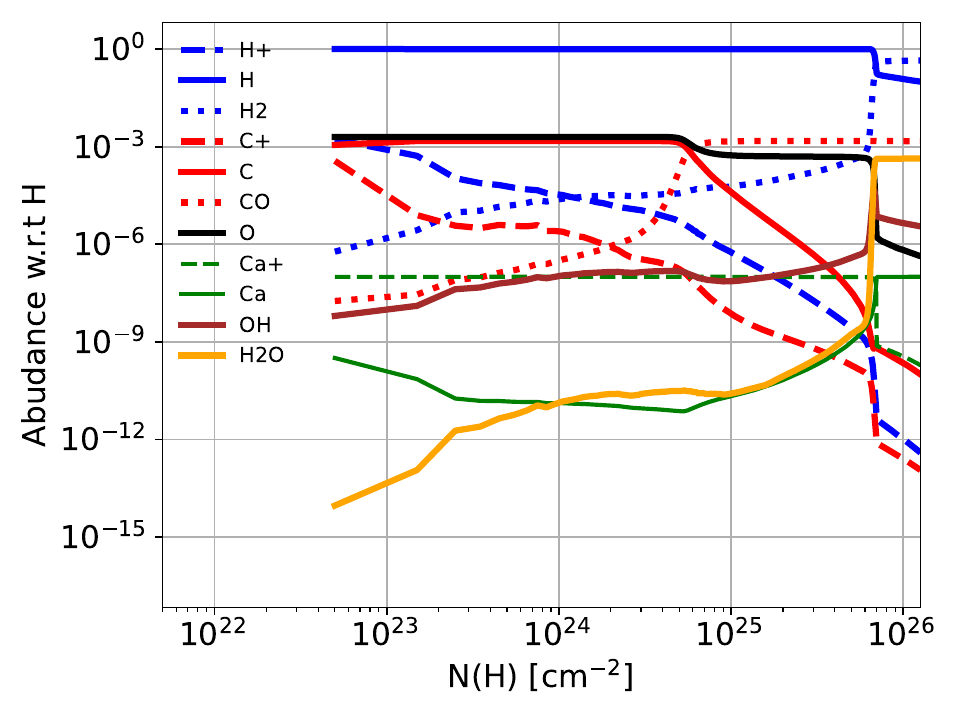}
\caption{Abundance for various species as function of the column density, {\it Left:} Model with $n_{\mathrm{H}} = 10^{12}$ cm$^{-3}$, $\log(U)=-3$, and $\Delta V = 250$ km s$^{-1}$, {\it Right:} Model with $n_{\mathrm{H}} = 10^{14}$ cm$^{-3}$, $\log(U)=-5$, and $\Delta V = 250$ km s$^{-1}$.}
\label{fig_depth}            
\end{figure*}  
\begin{figure*}[!ht]
\centering
\includegraphics[angle=0,width=6.0cm,height=4.5cm,clip]{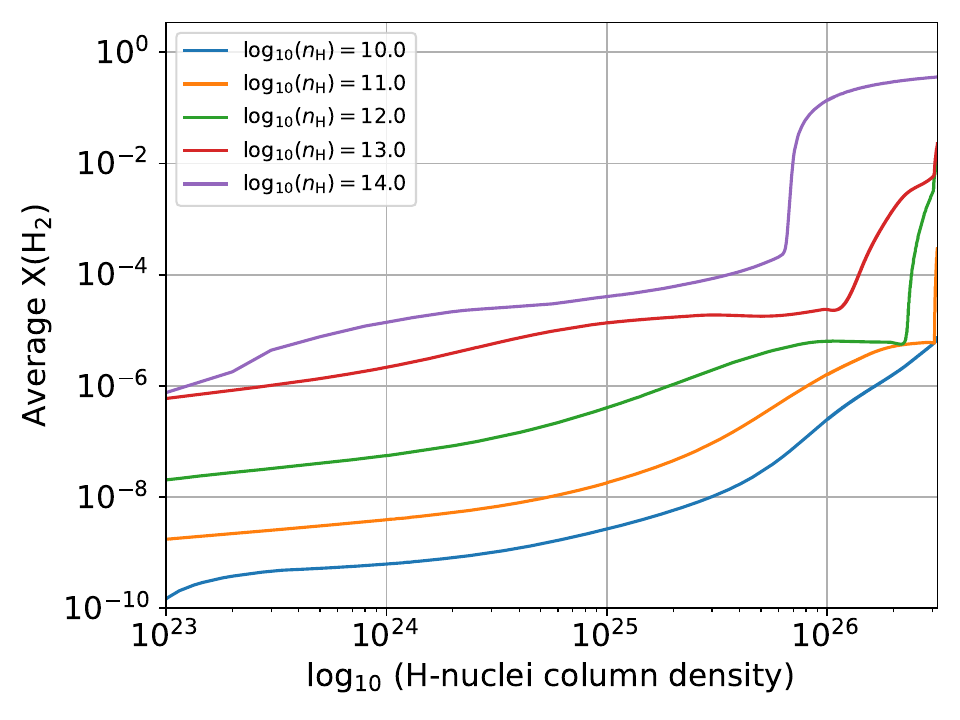}
\includegraphics[angle=0,width=6.0cm,height=4.5cm,clip]{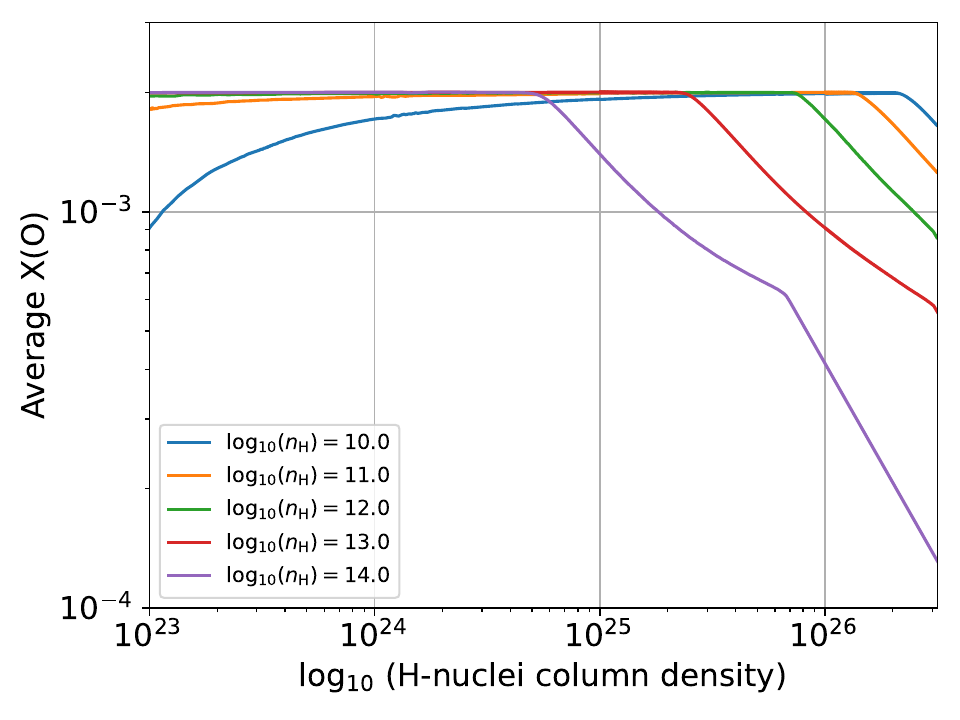}
\includegraphics[angle=0,width=6.0cm,height=4.5cm,clip]{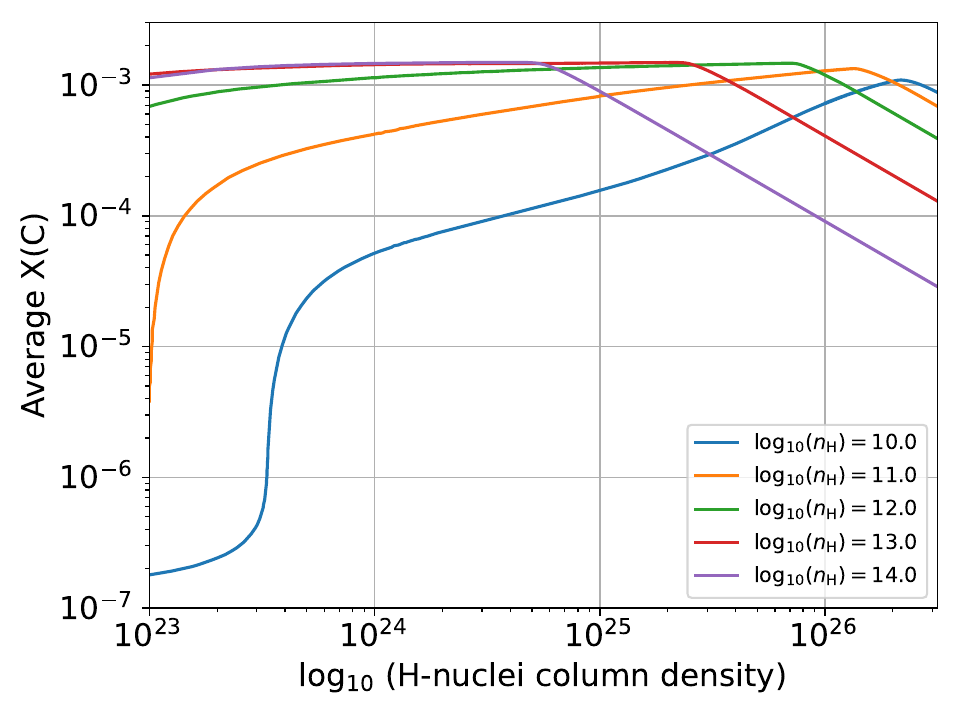}
\includegraphics[angle=0,width=6.0cm,height=4.5cm,clip]{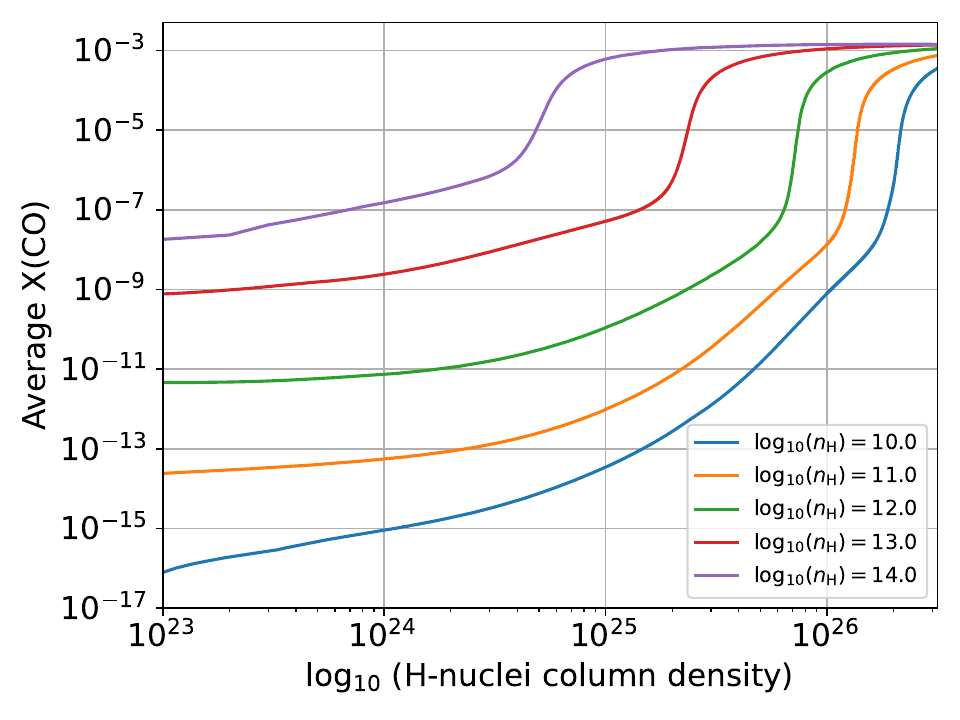}
\includegraphics[angle=0,width=6.0cm,height=4.5cm,clip]{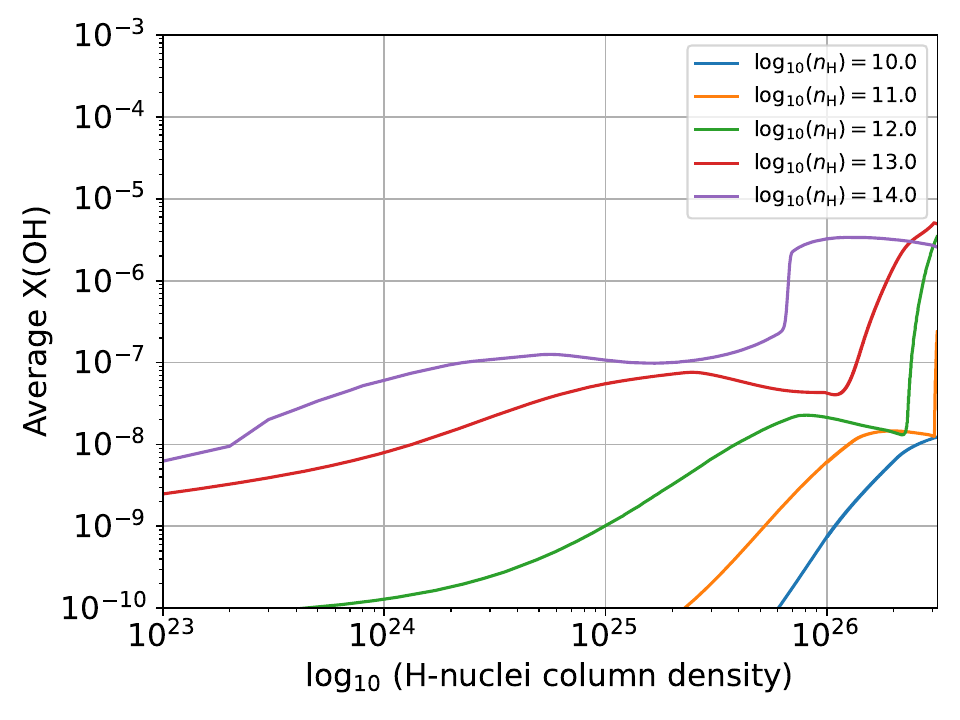}
\includegraphics[angle=0,width=6.0cm,height=4.5cm,clip]{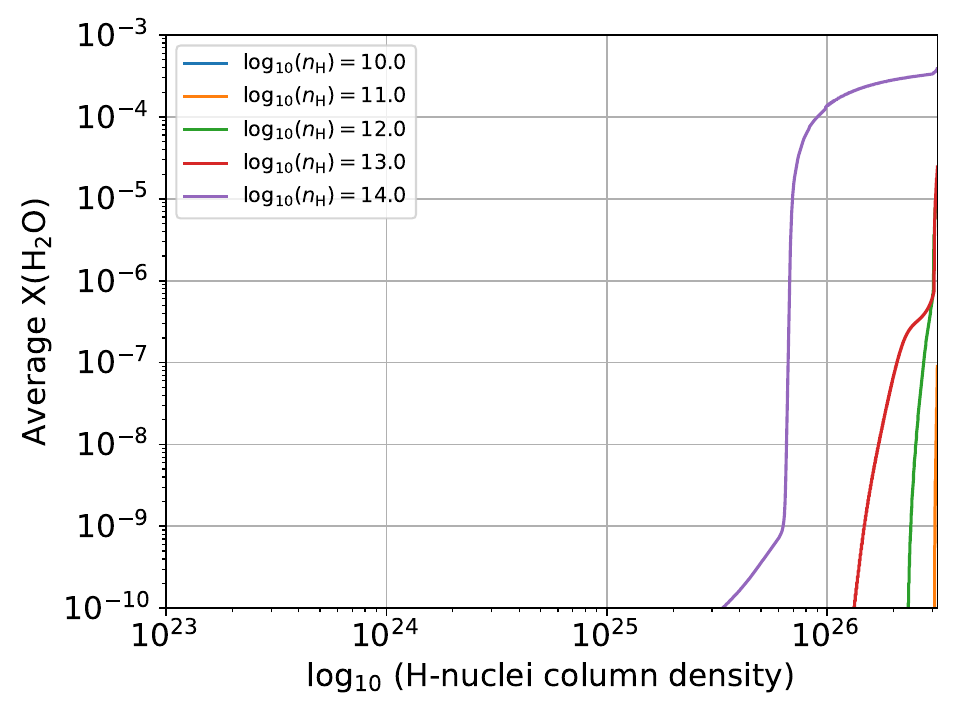}
\caption{Column density-averaged abundances for H$_2$, O, C, CO, OH, and H$_2$O. The total H-column density varies slightly between models as the grid is determined by the code.}
\label{fig_abundances}            
\end{figure*}  
\noindent
namely: $N(\mathrm{H)_{min}}$$\sim  $($10^{23}$--$10^{24}$)\,cm$^{-2}$  for
typical  SMBH  masses,  $M_{\mathrm{SMBH}}$, and  AGN  luminosities, $L_{\mathrm{AGN}}$. 
Furthermore, it is necessary for  the so-called  radiation-pressure-confined (RPC)  ionization layers  
to be established within the BLR$^{+}$s in a universal  fashion (in pressure
equilibrium with  deeper neutral BLR gas  layers) and to do so  across eight
orders of  magnitude of  AGN power \citep{Bas14,  Bas18}. Thus,  this value  must be
$N(\mathrm{H})$$\geq $$10^{24}$ cm$^{-2}$  \citep{Bas14}.    This result is
similar to  the $N(\mathrm{H)_{min}}$$\sim  $$10^{24.5}$\,cm$^{-2}$ value deduced
from   the   \ion{Ca}{ii}   triplet    line   observations,     tracing
low-ionization/neutral BLR gas layers \citep{fer89}. A maximum value
for the total $N$(BLR$^{+}$+BLR$^{0}$) column density of the various BLR
gas  structures can  then be    simply set as:  $N(\mathrm{H})_{\mathrm{max}}=N(\mathrm{H})_{\mathrm{BLR}^{+}}\left(1+M_{\mathrm{BLR}^{0}}/M_{\mathrm{BLR}^{+}}\right) \sim
N(\mathrm{H)_{min}} \times (10^2-10^4$).  For  a  $N(\mathrm{H)_{min}}=10^{23}$\,cm$^{-2}$,  we  adopted
$\log_{10}N(\mathrm{H)_{max}}=10^{26.5}$\,cm$^{-2}$,  with  a  $\log \Delta N(\mathrm{H})=0.5$ grid step\footnote{The exact computed column density is set by the code.}, and a  BLR density range of $n_\mathrm{H}=(10^{10},
10^{11}, 10^{12},  10^{13}, 10^{14})$\,cm$^{-3}$.   For the  ionization  parameter, $U,$ we  adopted  $U=0.1$
for $n_{\rm H}=10^{10}$\,cm$^{-3}$.  The other $U$-grid values are  chosen to make it so that
the ionizing photon density,  $n_{\gamma}=n_{\rm H}U$, remains invariant; this is required   by   the   AGN-induced  dust   sublimation condition holding
within   BLRs in all AGNs \citep{Bas14}.   We considered the metallicities of
$Z/Z_{\odot}$=5, a super-solar value that is typical
for BLR gas in galactic centers.  Finally we adopt a typical turbulent
linewidth  of  $\Delta   V$=$250\ $km\,s$^{-1}$  for  the  BLR  gas
\citep{Gas09}\footnote{This is an important parameter in any radiative
transfer model as it regulates the escape probability of
spectral line emission, gas cooling,  and thus gas thermodynamics.}.
Fine physical grids were chosen to improve the model convergence. 
The temperature iteration tolerance was set to $5 \times 10^{-3}$ (default value), while
the grid parameters were: hydrogen density, $n_{\mathrm H}$, ionization parameter, $U$,
and the total hydrogen column density, $N_{\mathrm H}$.

\section{Results and discussion}\label{results}

Figure~\ref{fig_depth} shows the abundance of various species as function of $N$(H) for the $n_{\rm H}=$10$^{12}$cm$^{-3}$, $\log(U)$=$-3$ model in the left panel and for the $n_{\rm H}=$  10$^{14}$cm$^{-3}$, $\log(U)$=$-5$ model in the right panel. From this figure, is apparent that O and C are the two neutral atoms with the highest abundances dominating throughout the BLR$^{0}$ column,  while deeper inside they react to form the dominant molecule expected in BLR$^{0}$'s, namely, CO. The latter can reach up to a relative abundance $X$(CO) $\sim 10^{-4}$--$10^{-3}$ deep inside that of BLR$^{0}$, whereas  $\rm H_2$ remains
only a trace element, as was already been established for dust-free BLRs by \citet{fer89} and \citet{Cro93}.
This is quite unlike  the ordinary ISM and, indeed, a CO-rich, yet H$_2$-poor gas phase is one of the peculiar hallmarks
of a dust-free, but hot and very dense gas chemistry. Evidence for exactly such chemistry has already been
uncovered for the hot, dust-free, inner gas disk around the B star \object{51 Oph} \citet{Thi05}, with CO
overtone lines even used to resolve the disk at sub-au scales \citet{grav5}.

From Fig.~\ref{fig_depth}, we see that for $n_{\rm H} =10^{12}$ cm$^{-3}$, the H$^{+}\rightarrow$H
phase transition occurs at $N$(H)$\sim 10^{21}$ cm$^{-2}$, with neutral atomic H remaining dominant throughout
the remaining BLR$^{0}$ column. Only at the highest gas densities ($n_{\rm H}=10^{14}$ cm$^{-3}$) and largest
BLR$^{0}$ columns ($N$(H) $\sim 5\times 10^{25}$ cm$^{-2}$) does the neutral atomic H becomes molecular (Fig.~\ref{fig_depth}).
 Atomic oxygen, with its ionization potential (IP) of 13.618 eV only slightly above that of H (IP=13.598 eV), 
 can remain neutral nearly throughout the H-rich BLR$^{0}$ column. Atomic carbon on the other hand,
with an IP of $\sim $11.26 eV, can remain (singly) ionized inside the  BLR$^{0}$ past the
H$^{+} \rightarrow$ H transition, but by $N$(H)$\geq 3 \times 10^{22}$ cm$^{-3}$ (and $n_{\rm H}$=$10^{12}$ cm$^{-3}$),
it becomes neutral. For higher densities though ($n_{\rm H}$=$10^{14}$ cm$^{-3}$) neutral atomic carbon dominates already from the BLR$^{0}$ surface. Regarding CO, for $n_{\rm H}$=10$^{12}$ cm$^{-3}$, its abundance reaches $X$(CO)$\sim$~10$^{-4}$ only when $N$(H)$\geq $$8\times 10^{25}$ cm$^{-3}$ is reached; however,
this transition happens in a much $''$shallower$''$  segment within the  BLR$^{0}$ column, $N$(H) $\sim$ $6\times 10^{24}$ cm$^{-3}$, at higher densities ($n_{\rm H}$=$10^{14}$ cm$^{-3}$). 

It expected from the ionization potentials  of Ca and  Ca$^+$ of IP=6.11\,eV  and 11.87\,eV, respectively, that these two species are expected to be abundant in BLR$^{0}$ \citep{fer89}. Figures~\ref{fig_depth} and \ref{fig_Ca_depth} show that the Ca\ion{II}\ abundance remains nearly constant across the  entire BLR$^{0}$  for  $n_{\rm H}$=$10^{12}$ cm$^{-3}$, and  it is only for  the high-density   model   ($n$=$10^{14}$ cm$^{-3}$) and large column densities ($\geq 6\times 10^{25}$ cm$^{-3}$) that its abundance drops sharply  due to e$^{-}$  recombination into Ca; meanwhile, Ca-bearing molecules have very low abundances in all models. Only depletion onto dust  grains has the ability to change the gas-phase elemental abundance of calcium, but this is not an option in dustless BLRs.

Figure~\ref{fig_abundances} shows the $N$(H)-averaged abundances of H$_2$, O, C, CO, OH, and H$_2$O. The attenuation of the ionizing and molecule-dissociating radiation results in neutral and molecular species becoming abundant in BLR$^{0}$. Nevertheless H$_2$ itself remains only a trace species, except at the highest densities as well as the highest column densities, where $X$(H$_2$) $\sim 0.1$ (Fig.~\ref{fig_abundances}, upper left). At high gas densities and temperatures up to $\sim$1000\,K: e$^{-}$ recombination, neutral-neutral reactions with activation energy, and (potentially) three-body reactions are rapid because photoionization rates are $\propto $$n_{\rm H}$, while recombination rates  are $\propto n_{\rm H}^2$. These reactions can overcome the (neutral-atom and molecule)-destructive reactions such as photoionization, collisional ionization, photodissociation, and the collisional molecular dissociation. The OH abundance is higher than that of H$_2$O in the $n_{\rm H} = 10^{12}$ cm$^{-3}$ model, but it remains a secondary O-carrier; whereas for the highest density model ($n_{\rm H} = 10^{14}$ cm$^{-3}$) and at $N$(H), water is the second most abundant oxygen carrier.

The chemical  molecule-formation path in dense and hot gas starts with H$_2$ formation in the presence of sufficient C$^+$, C, and O. In the absence of dust grains, H$_2$ is then formed solely via gas-phase reactions \citep{Cro93}. Once formed, H$_2$ reacts with atomic oxygen to form OH, which in turn can react with H$_2$ to form H$_2$O, or with singly ionized carbon to form CO via the reaction C$^+$ + OH $\rightarrow$ CO$^+$, followed by CO$^+$ + H $\rightarrow$ CO + H$^+$ \citep{Elitzur 1979, Hollenbach1989, ThiBik2005}. More details are given in Appendix~\ref{formation_destruction}.  Alternatively, OH can react with C to form directly CO at very high densities. In the absence of dust grains, atomic C, H$_2$, and CO molecules are  shielded against photodissociation solely via line absorption, although the large intrinsic turbulence width significantly decreases  the efficacy thereof. The CO/H$_2$ ratio is much higher than in the ordinary dusty  molecular ISM, making CO the most abundant molecule and also making its lines potentially good   BLR$^{0}$ tracers. Furthermore, atomic oxygen, atomic carbon, OH, H$_2$O, and CO, once formed, act as very efficient gas coolants. This provides the capacity to strongly impact the gas thermal states, decreasing their temperature in deeper BLR$^{0}$ layers. In principle this can push gas temperatures below the dust sublimation limit ($\sim $1000-1500\,K, see Appendix~\ref{appendx_gas_temperature}), allowing some dust to survive. Even trace amounts of dust surviving under BLR$^{0}$ conditions can have a very strong impact on H$_2$ formation and to  make the BLR$^{0}$
very H$_2$-rich (\citet{Cro93}). When very few dust grains survive in BLR$^{0}$, the enhanced H$_2$ formation
is  due to the dust-enhanced attenuation of the molecule-dissociating AGN radiation field rather than to the
classical H$_2$ formation route on dust grains. Indeed, even at $T_{\rm dust}$$\sim $600~K, the grains are still too warm to allow efficient formation of H$_2$ on their surfaces \citep{Thi2020}, while even a little dust can strongly
attenuate the photodissociating far-UV flux, enhancing  molecule formation. At the same time, gas-grain collisions, along with molecules, will cool the gas rapidly (especially at high gas densities), enhancing the conditions for a greater scale of molecule formation. Finally, H$_2$ can also form on PAHs \citep{Boschman2015A}, which can survive in the hot environments of BLRs where silicate dust grains cannot. Thus inclusion of dust grains and PAHs in BLRs  can further enhance the abundance of the molecular species such as H$_2$ and CO in their BLR$^{0}$. Here, we must also note that for the inwards increasing gas density profiles expected in radiation pressure confined (RPC) BLR models \citep{Bas14}, the neutral atomic and molecular species abundances within BLR$^{0}$'s can only be higher than those found for our uniform-density  models.

Finally, for the strongly UV-deficient AGN SEDs associated with low $(dM/dt)/M^2_{\rm SMBH}$ ratios (i.e., low accretion rates and/or hypermassive black holes, \citealt{Lade11}), their BLRs are expected to be even more neutral-atom- and molecule-rich,  perhaps even completely lacking a BLR$^{+}$ as an outside layer of the BLR$^{0}$ (yielding the so-called "true" type 1 AGNs). These cases, along with the effects of small amounts of dust and PAHs and assuming a disk BLR configuration (as indicated by high-resolution BLR observations \citet{Stur18})  will be the subject of a forthcoming paper.

\subsection{ BLR$^{0}$ lines: A new spectral window onto AGNs and extreme gas velocity fields near SMBHs}

Atomic oxygen can emit a series of lines in the UV/optical/NIR and far-infrared (FIR) wavelength regimes (\citet{Bh95} their Table 2).
From these,  we can see that only its two fine structure lines at 63.17\,$\mu $m and 145.5\,$\mu $m (due to $\vec{L}\cdot\vec{S}/r^3$-type interaction term)
can emanate from the BLR$^{0}$, where the prevailing conditions are adequate for their collisions excitation up to full
LTE levels. The rest of the \ion{O}{I} lines, given the energy transitions between the ($n$, $J$) principal quantum numbers, are energetically too demanding to be collisionally excited within BLR$^{0}$'s. Pumping by Ly$\beta$ photons is necessary for the NIR \ion{O}{I} lines to be excited, placing them in the so-called low-ionization regions of the BLR$^{+}$ (where such photons are readily available). \citet{Dias2023} has recently detected a very wide, double peak, \ion{O}{I} NIR line ($\lambda$11297) in  \object{Zw 002} (Seyfert I), which may indeed
be marking the low-ionization layer of a BLR$^{+}$. The two fine structure lines of  \ion{C}{I} at $\sim $492\,GHz and $\sim $809\,GHz,
along with the $J$-rotational lines of CO with $J$=1--0 (115\,GHz) up to  $J$=6-5, 7-6 (691\,GHz, 806\,GHz),
are accessible through the Earth's atmosphere and will be LTE-excited in a BLR$^{0}$. Deep ALMA observations; while, they cannot resolve  \ion{C}{I} and CO line emission in a BLR$^{0}$, they are able to identify broad BLR$^{0}$ lines provided  that multi-tuned ALMA correlator setups are used to cover the $\sim$(few)$\times $($10^{3}$--$10^{4}$)\,km\,s$^{-1}$ velocity range. The highest-$J$ CO lines  accessible through the atmosphere (i.e., $J$=6-5, 6-7) are the best choices since for  LTE-excitation:
$\int_{\Delta V} S^{({\rm line})} _{\nu} d\nu\propto (J+1)^3\nu ^{3} _{\circ} T_{\rm kin}$\footnote{Slightly redshifted AGNs can
be used to place such lines into much better atmospheric windows, without  sacrificing too much in terms of the received flux.}
($\nu _{\circ}$$\sim $115\,GHz). High-resolution ALMA observations of high-$J$ CO lines of nearby AGNs do exist, (\citealt{gal16}),
but their limited velocity coverage ($\sim $600\,km\,s$^{-1}$), precludes them from identifying any ultra-wide BLR$^{0}$ CO line emission. Warm CO-rich gas can emit ro-vibrational lines in the NIR-MIR that are then accessible to spatially resolved observations with IR interferometers, such as GRAVITY+ or MATISSE at ESO. GRAVITY has already been used in the past to identify dustless hot and dense CO-rich gas around \object{51 Oph}. Although water can be abundant in the densest models, water lines are difficult to detect from ground-based telescopes and also require a favorable redshift.

\subsection{Spectral line illumination of SMBHs: Possible new GR tests}
Depending on their actual luminosities,  the various neutral-atomic and/or molecular lines from
 the BLR$^{0}$ gas disk can provide a novel spectral line, rather than a continuum, background
 illumination of the SMBHs in  galactic centers. This aspect can enable new tests of Kerr versus non-Kerr Spacetimes  around these ultra-relativistic objects, as recently proposed by \citet{kos24}.

\section{Conclusions}

Our work demonstrates the likely presence of neutral atoms and CO molecules in a 
potentially massive neutral gas BLR$^{0}$ phase underlying the thin ionized layers of BLR$^{+}$ in AGNs. In the dustless, albeit warm and dense, BLR$^{0}$ gas, the prevailing molecule is CO; whereas  H$_2$  remains only a trace species. The fine structure lines of neutral atomic oxygen and carbon at FIR and submm wavelengths can uniquely trace the  BLR$^{0}$ gas mass, while CO, with its rotational and ro-vibrational CO lines at mm/submm and NIR lines (also expected to be bright) open up a new observational spectral window to BLRs in AGNs and the extreme velocity fields near SMBHs.
Gas density gradients, minute amounts of surviving dust (especially the thermally-resistant PAHs), and UV-photon deficient AGN SEDs (the result of low-accretion and/or hypermassive BHs in galactic centers) are all expected to further boost the abundances of neutral atoms and molecules in BLR$^{0}$ and will be  studied in detail in a forthcoming paper.

\begin{acknowledgements}
We would like to offer thanks to the  referee, that helped us
clarify some finer, yet important, points in the presented work. PPP would like to acknowledge
the hospitality of his colleague Wing-Fai Thi, in his apartment in Munich, where these ideas were  
developed.
\end{acknowledgements}

%
%

\begin{appendix}

\section{Three-body reactions}\label{Appendix_chem}

At densities $>$~10$^{10}$ cm$^{-3}$, the rate of three-body reactions becomes competitive with two-body reactions. Three-body formation and destruction reactions of H$_2$  are accounted by default by the code. We complemented the reaction networks with three-body reactions taken from the UMIST2012 \citep{McElroy2013}. 

\begin{table}[!ht]
    \centering
    \caption{Three-body (collider) reactions added to the standard {\sc Cloudy} network. The rate coefficients are taken from \citep{McElroy2013}. M designates H or H$_2$.}
    \begin{tabular}{lcl}
    \hline
        &\textbf{Reaction}&\\
    \hline
    CO + H + M & $\rightarrow$ &  O + C + H + M\\
    CO + H$_2$ + M & $\rightarrow$ & O + C + H$_2$ + M\\
    CO + H + HOC$^+$ + M & $\rightarrow$ & HCO$^+$ + CO + H + M\\
    CO + H$_2$ + HOC$^+$ + M & $\rightarrow$ & HCO$^+$ + CO + H$_2$ + M\\
    C + O + H + M & $\rightarrow$ & CO + H + M\\
    C + O + H$_2$ + M & $\rightarrow$ & CO + H$_2$ + M\\
    H + O + H + M & $\rightarrow$ & OH + H + M\\ 
    H + O + H$_2$ + M & $\rightarrow$ & OH + H$_2$ + M\\
    H + OH + H + M & $\rightarrow$ & H$_2$O + H + M\\
    H + OH + H$_2$ + M & $\rightarrow$ & H$_2$O + H$_2$ + M\\
    H + CH$_3$ + H + M & $\rightarrow$ & CH$_4$ + H + M\\
    H + CH$_3$ + H$_2$ + M & $\rightarrow$ & CH$_4$ + H$_2$ + M\\
    H + NH$_2$ + H + M & $\rightarrow$ & NH$_3$ + H + M\\
    H + NH$_2$ + H$_2$ + M & $\rightarrow$ & NH$_3$ + H$_2$ + M\\
    H$_2$ + N + H + M & $\rightarrow$ & NH$_2$ + H + M\\
    H$_2$ + N + H$_2$ + M & $\rightarrow$ & NH$_2$ + H$_2$ + M\\
    O + O + H + M & $\rightarrow$ & O$_2$ + H + M\\
    O + O + H$_2$ + M & $\rightarrow$ & O$_2$ + H$_2$ + M\\
    H$_2$ + C + H + M & $\rightarrow$ & CH$_2$ + H + M\\
    H$_2$ + C + H$_2$ + M & $\rightarrow$ & CH$_2$ + H$_2$ + M\\
    H$_2$ + CH + H + M & $\rightarrow$ & CH$_3$ + H + M\\
    H$_2$ + CH + H$_2$ + M & $\rightarrow$ & CH$_3$ + H$_2$ + M\\
    C + N + H + M & $\rightarrow$ & CN + H + M\\
    C + N + H$_2$ + M & $\rightarrow$ & CN + H$_2$ + M\\
    C$^+$ + O + H + M & $\rightarrow$ & CO$^+$ + H + M\\
    C$^+$ + O + H$_2$ + M & $\rightarrow$ & CO$^+$ + H$_2$ + M\\
    C + O$^+$ + H + M & $\rightarrow$ & CO$^+$ + H + M\\
    C + O$^+$ + H$_2$+ M & $\rightarrow$ & CO$^+$ + H$_2$ + M\\
    \hline
    \end{tabular}
    \label{tab:reactions}
\end{table}

\section{Additional reaction}\label{additional_reaction}

The reaction H + CO has two branches: H + C + O and C + OH. For completeness, we also added the reaction H + CO $\rightarrow$ C + OH, which has an activation energy of 77700~K, to the {\sc CLOUDY} network. In our models, the H + C + O outcome dominates over the formation of OH.

\section{Calcium chemistry}\label{calcium_network}

\begin{figure*}[!h]
\centering
\includegraphics[angle=0,width=9cm,height=6cm,clip]{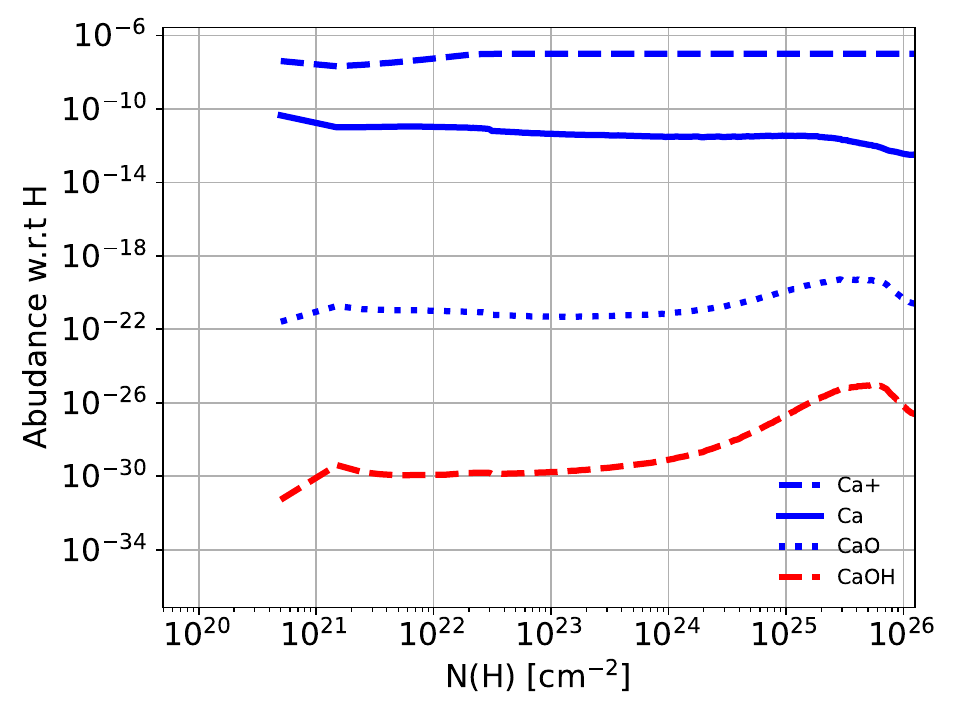}
\includegraphics[angle=0,width=9cm,height=6cm,clip]{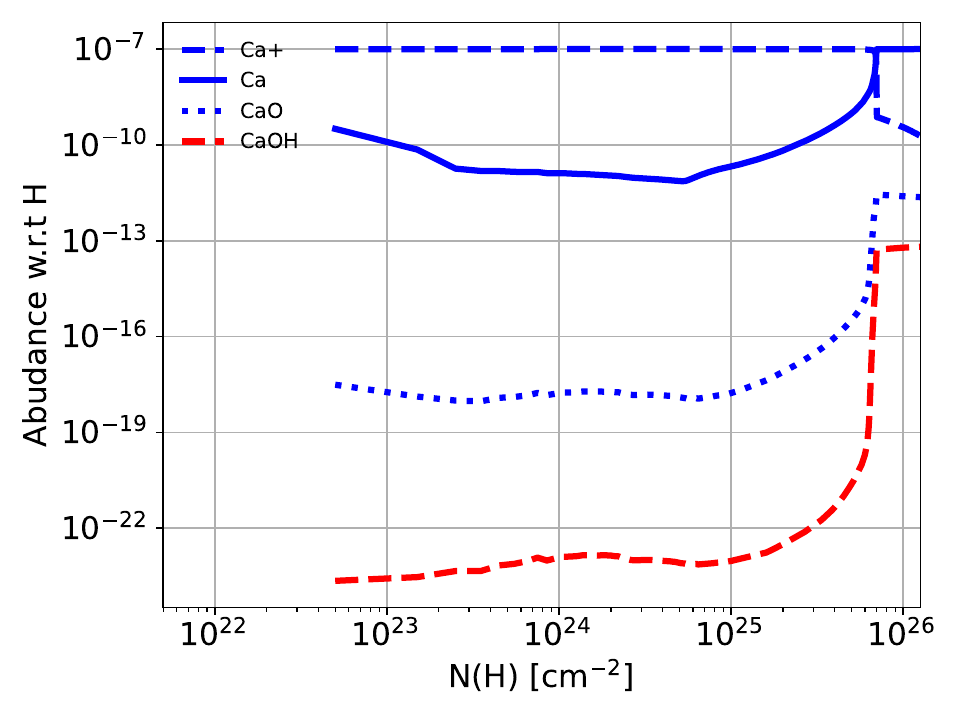}
\caption{Abundance for Ca-bearing species as function of the column density, {\it Left:} A model with $n_{\mathrm{H}} = 10^{12}$ cm$^{-3}$, $\log(U)=-3$, $\Delta V = 250$ km s$^{-1}$, {\it Right:} A model with $n_{\mathrm{H}} = 10^{14}$ cm$^{-3}$, $\log(U)=-5$, $\Delta V = 250$ km s$^{-1}$.}
\label{fig_Ca_depth}            
\end{figure*}

{\sc Cloudy} accounts by default for the ionization and recombination of atomic and ionized Ca. We complemented the default network with chemical reactions that includes Ca-bearing molecules. The reactions and rates are taken from the UMIST2022 network \citep{Millar2024}. The abundance for the Ca-bearing species for the two typical models are shown in Fig.~\ref{fig_Ca_depth}. The molecular Ca-species are at most two orders of magnitude less abundant than Ca or Ca$^+$. In the high density model ($n_{\rm H} = 10^{14}$ cm$^{-3}$), atomic Ca is the most abundance Ca-species so that the Ca\ion{II}\ lines probe only a fraction of the total neutral gas at high column densities.

\begin{table}[!ht]
    \centering
    \caption{Ca species reactions added to the standard {\sc Cloudy} network. The rate coefficients are taken from \citep{Millar2024}. Reactios with CRPHOT are Cosmic-Ray induced photoionization reactions and reactions with PHOTON are Far-UV photodissociation reactions.}
    \begin{tabular}{lcl}
    \hline
        &\textbf{Reaction}&\\
    \hline
Ca + CRPHOT & $\rightarrow$ &  Ca$^+$ + e$^-$\\
CaO  + CRPHOT & $\rightarrow$ & Ca  + O\\
CaOH  + CRPHOT & $\rightarrow$ & Ca  + OH\\
Ca  + OH & $\rightarrow$ & CaO  + H\\
Ca  + H$_2$O & $\rightarrow$ & CaO  + H$_2$\\
Ca  + H$_2$O & $\rightarrow$ & CaOH  + H\\
Ca  + CO$_2$ & $\rightarrow$ & CaO  + CO\\
Ca  + O$_2$ & $\rightarrow$ & CaO  + O\\
CaO  + H & $\rightarrow$ & Ca  + OH\\
CaO  + H$_2$ & $\rightarrow$ & Ca  + H$_2$O\\
CaO  + O & $\rightarrow$ & Ca  + O$_2$\\
CaO  + CO & $\rightarrow$ & Ca  + CO$_2$\\
CaOH  + H & $\rightarrow$ & Ca  + H$_2$O\\
CaOH  + H & $\rightarrow$ & CaO  + H$_2$\\
CaO  + PHOTON & $\rightarrow$ & Ca  + O\\
CaOH  + PHOTON & $\rightarrow$ & Ca  + OH\\
    \hline
    \end{tabular}
    \label{tab:Ca_reactions}
\end{table}

\section{Photodissociation and self-shielding}\label{photodissociation}

\begin{figure}[!h]
\centering
\includegraphics[angle=0,width=9cm,height=6cm,clip]{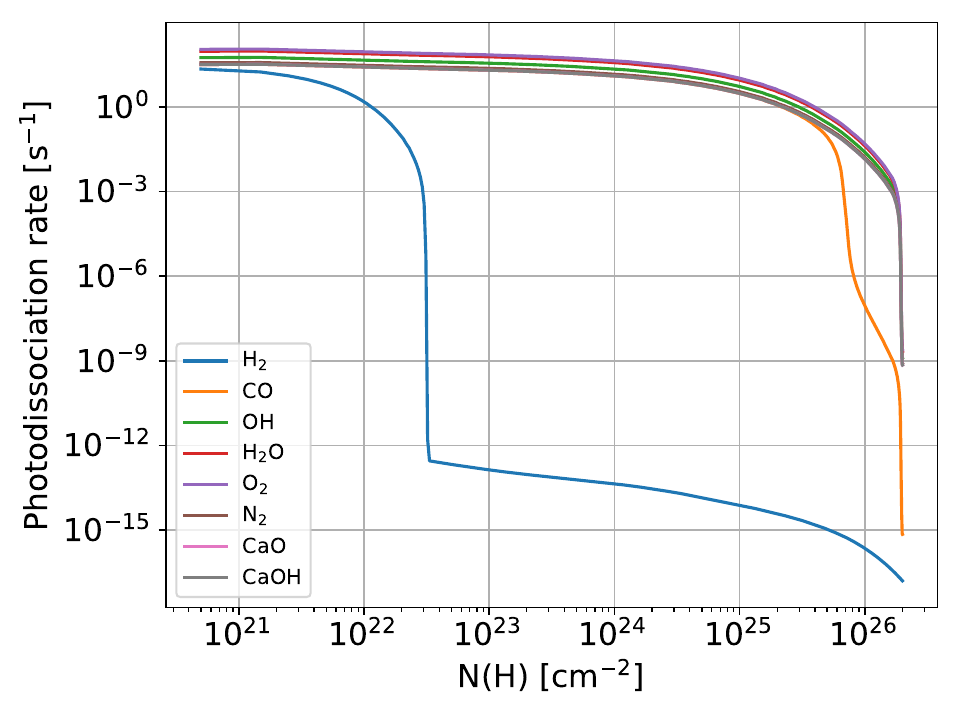}
\caption{Gas temperature profile for a model with $n_{\mathrm{H}} = 10^{12}$ cm$^{-3}$, $\log(U)=-3$, $\Delta V = 250$ km s$^{-1}$.}
\label{fig_photodissociation_rates}            
\end{figure}

Self-shielding against photodissociation for H$_2$ and CO are accounted 
for by the code as illustrated by Fig.~\ref{fig_photodissociation_rates}.
In the figure, the photodissociation rates for H$_2$ and CO drop suddenly as soon as the relevant threshold column density is reached.

\section{The BLR gas temperature profile}
\label{appendx_gas_temperature}  

\begin{figure}[!h]
\centering
\includegraphics[angle=0,width=9cm,height=6cm,clip]{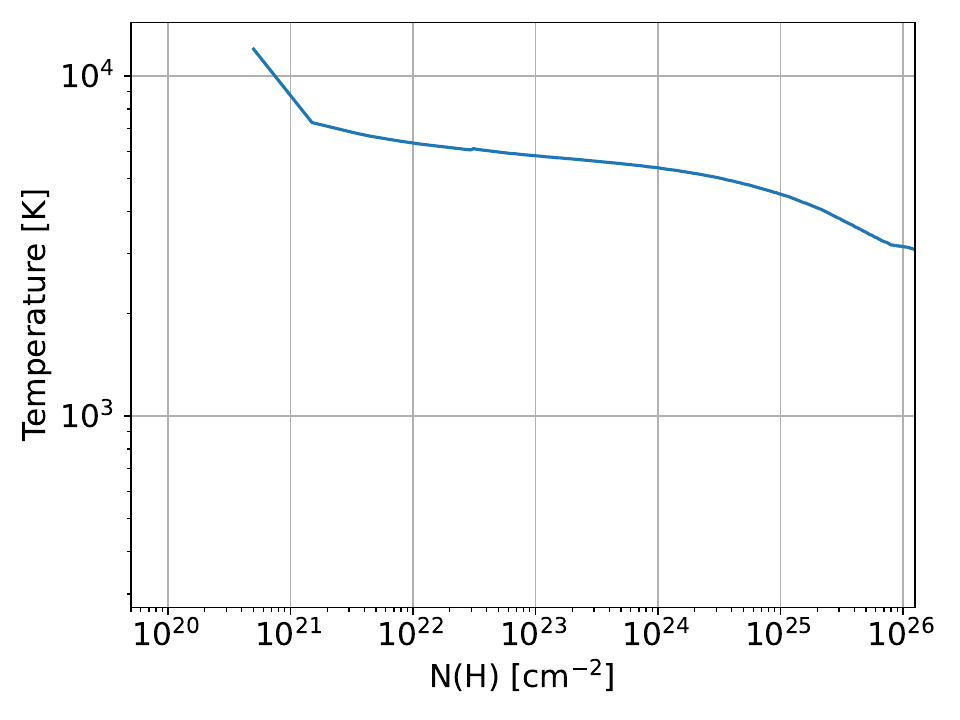}
\caption{Gas temperature profile for a model with $n_{\mathrm{H}} = 10^{12}$ cm$^{-3}$, $\log(U)=-3$, $\Delta V = 250$ km s$^{-1}$.}\label{fig_temperature1}
\includegraphics[angle=0,width=9cm,height=6cm,clip]{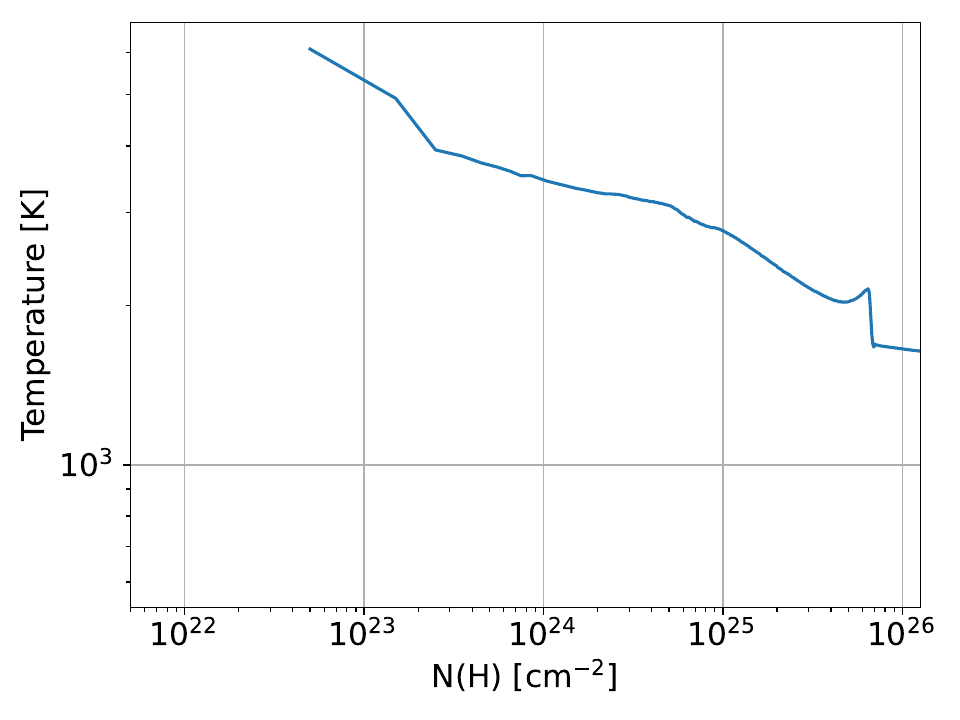}
\caption{Gas temperature profile for a model with $n_{\mathrm{H}} = 10^{14}$ cm$^{-3}$, $\log(U)=-5$, $\Delta V = 250$ km s$^{-1}$.}\label{fig_temperature2}          
\end{figure}

The gas temperature as function of depth for the 10$^{12}$ and 10$^{14}$ cm$^{-3}$ models are shown in Fig.~\ref{fig_temperature1} and Fig.~\ref{fig_temperature2}. The gas temperatures decrease with depth as photons are absorbed and cooler and molecular gas are more efficient gas coolants.

\section{H$_2$ and CO main formation and destruction}\label{formation_destruction}

\begin{figure*}[!ht]
\centering
\includegraphics[angle=0,width=9cm,height=6cm,clip]{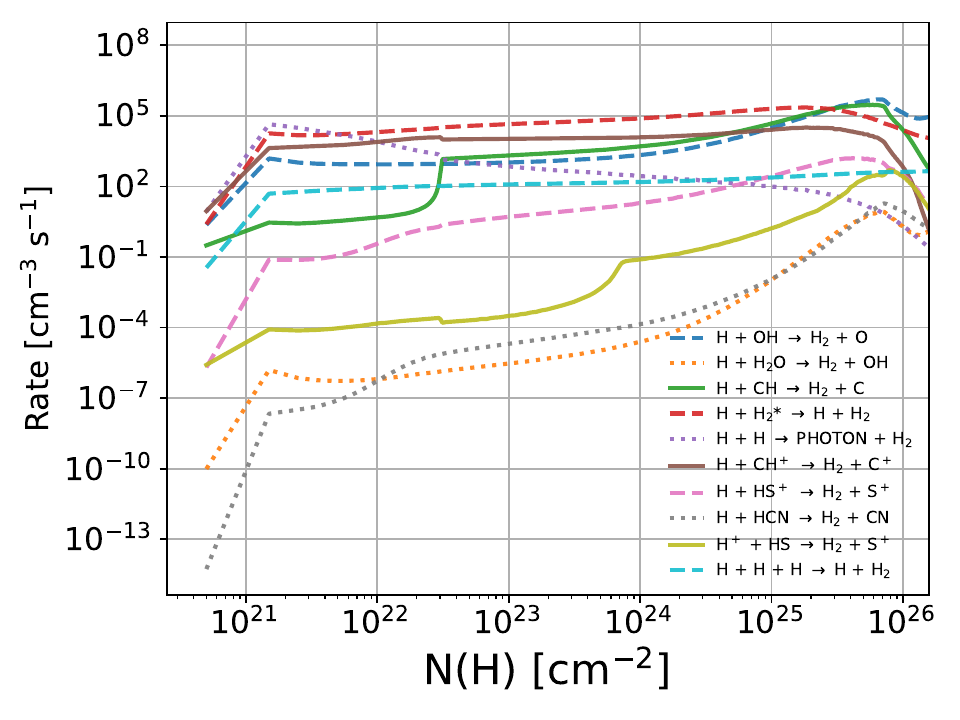}
\includegraphics[angle=0,width=9cm,height=6cm,clip]{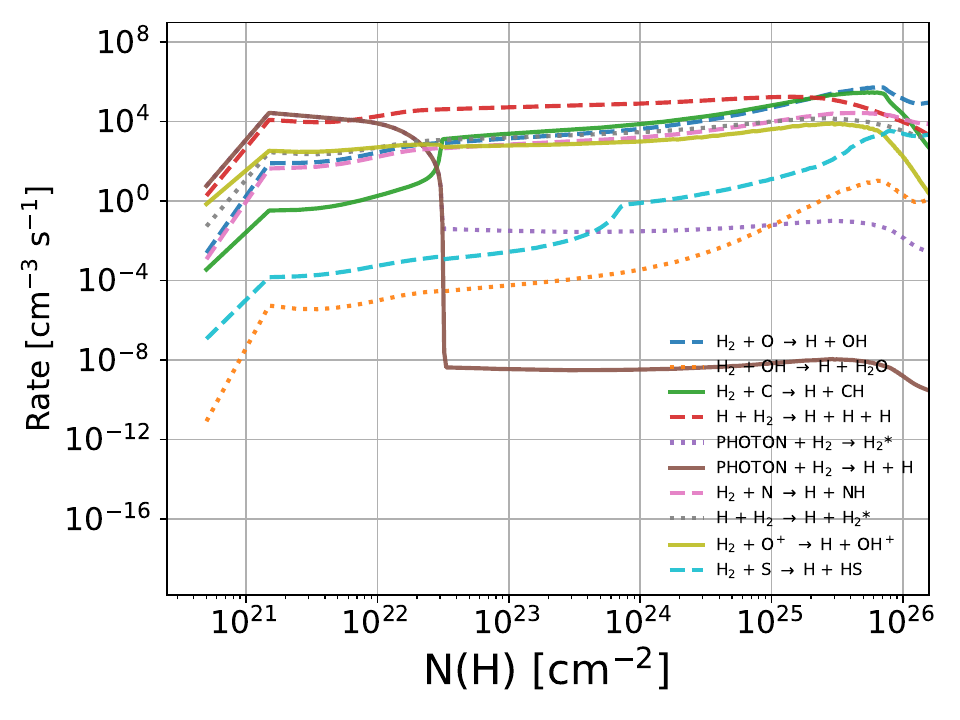}
\includegraphics[angle=0,width=9cm,height=6cm,clip]{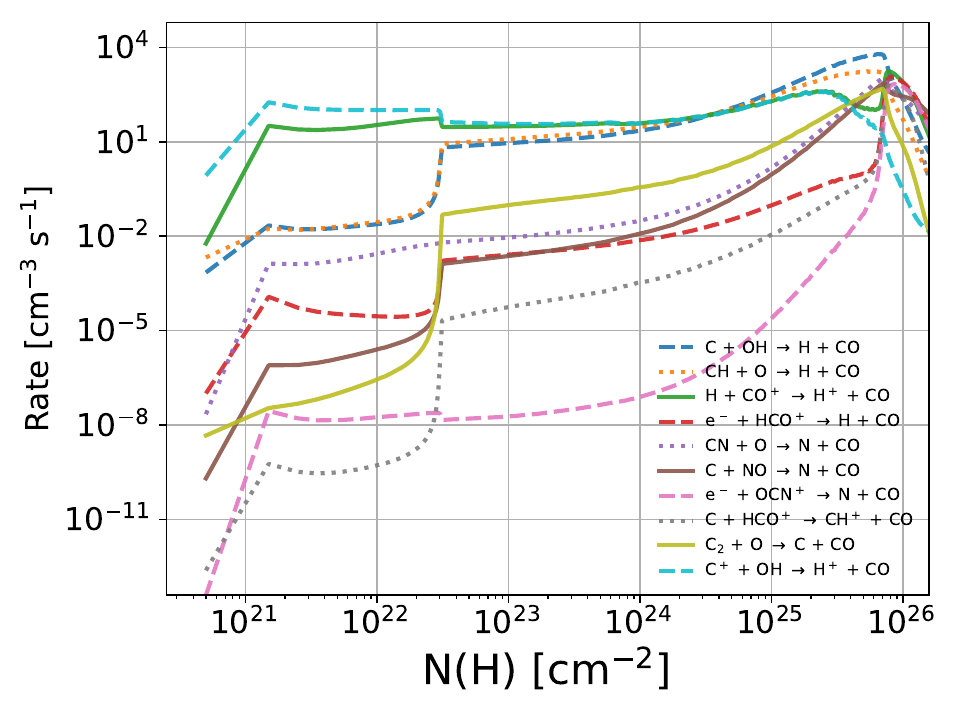}
\includegraphics[angle=0,width=9cm,height=6cm,clip]{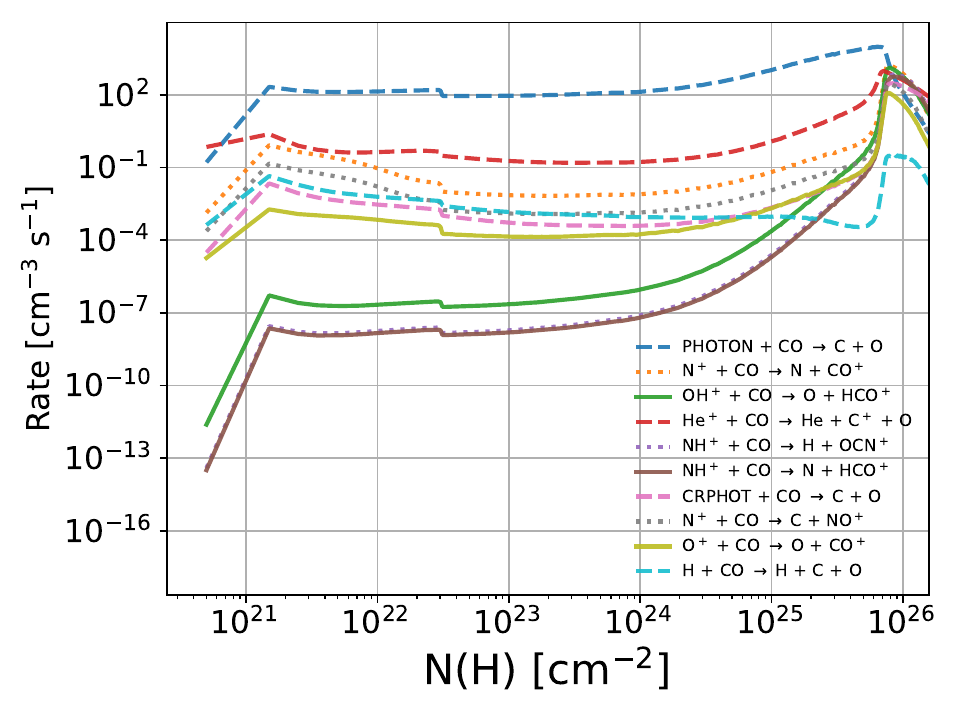}
\caption{Main formation (left panels) and destruction (right panels) reactions for H$_2$ and CO for the $n_{\mathrm{H}} = 10^{12}$ cm$^{-3}$, $\log(U)=-3$, $\Delta V = 250$ km s$^{-1}$ model. Only the top 10 main formation and destruction reactions are shown}.
\label{fig_formation_destruction}            
\end{figure*}  
Figure~\ref{fig_formation_destruction} illustrates the main formation and destruction routes for H$_2$ and CO at steady-state for a model with density of 10$^{12}$ cm$^{-3}$. The main reaction that actually forms H$_2$ from simpler H atoms is H$_2$ radiative recombination reaction. H$_2$ is destroyed at high column density by the formation of OH, which can react with C to form CO. OH is formed by the reaction O + H$_2$  with an activation barrier of $\sim$~4000~K. This reaction is only relevant for gas hotter than $\sim$~1000~K. The current network does not account for vibrational-excited H$_2$ reactions. OH can react with H to reform H$_2$. At low column density CO is mostly destroyed by photodissociation. At column densities greater than 10$^{26}$ cm$^{-2}$ various reactions involving ions (N$^+$, OH$^+$, and He$+$ contribute to CO destruction. A detailed discussion of the reaction paths can be found in \cite{ThiBik2005}. The H$_2$ formation via the radiative association H + H is very low and its rate depends on the excitation state of H \citep{Latter1991}. In our models, the radiative association reaction is only important at very low column densities.
\end{appendix}
  
\end{document}